\documentclass[twocolumn,showpacs,preprintnumbers,amsmath,amssymb]{revtex4}
\usepackage{amsmath}
\usepackage{bm}
\usepackage{epsfig}
\usepackage[colorlinks=true,linkcolor=blue]{hyperref}

\newcommand{\be}{\begin{equation}}
\newcommand{\ee}{\end{equation}}

\newcommand{\bea}{\begin{eqnarray}}
\newcommand{\eea}{\end{eqnarray}}

\newcommand{\al}{\alpha}

\newcommand{\gm}{\gamma}

\newcommand{\dl}{\delta}
\newcommand{\Dl}{\Delta}
\newcommand{\eps}{\epsilon}

\newcommand{\et}{\eta}

\newcommand{\kp}{\kappa}
\newcommand{\lm}{\lambda}
\newcommand{\Lm}{\Lambda}

\newcommand{\rh}{\rho}

\newcommand{\sg}{\sigma}

\newcommand{\ta}{\tau}

\newcommand{\ph}{\phi}

\newcommand{\om}{\omega}

\newcommand{\rarrow}{\rightarrow}

\newcommand{\nn}{\nonumber}

\begin{document}

\title{Magnetohydrodynamics and Plasma Cosmology}

\author{Kostas Kleidis$^{1,2}$, Apostolos Kuiroukidis$^{1,3}$, Demetrios Papadopoulos$^1$
and Loukas Vlahos$^1$}

\affiliation{$^1$Department of Physics, Aristotle University of
Thessaloniki, 54124 Thessaloniki, Greece}

\affiliation{$^2$Department of Civil Engineering, Technological
Education Institute of Serres, 62124 Serres, Greece}

\affiliation{$^3$Department of Informatics, Technological
Education Institute of Serres, 62124 Serres, Greece}

\date{\today}

\begin{abstract}

We study the linear magnetohydrodynamic (MHD) equations, both in
the Newtonian and the general-relativistic limit, as regards a
viscous magnetized fluid of finite conductivity and discuss
instability criteria. In addition, we explore the excitation of
cosmological perturbations in anisotropic spacetimes, in the
presence of an ambient magnetic field. Acoustic, electromagnetic
(e/m) and fast-magnetosonic modes, propagating normal to the
magnetic field, can be excited, resulting in several implications
of cosmological significance.

\end{abstract}


\maketitle

\section{Introduction}

Although Plasma Physics and Cosmology are two well-established
fields of Theoretical Physics, the formulation of
magnetohydrodynamics in curved spacetime is a relatively new
development~\cite{1}. In particular, in spite the fact that MHD
processes in flat spacetime gained much attention~\cite{2}, only
recently we were able to discuss exact spherically symmetric MHD
solutions within the context of General Relativity (GR)~\cite{3},
something that gave rise to efforts of exploring MHD processes in
the vicinity of central engines~\cite{4}, ~\cite{5}, ~\cite{6}. On
the other hand, magnetic fields are known to have a widespread
presence in our Universe, being a common property of the
intergalactic medium in galaxy clusters~\cite{7}, while, reports
on Faraday rotation imply magnetic fields of significant strength
at high redshifts~\cite{8}, ~\cite{9}. These results lead the
scientists to go even further and to look for gravitational
instabilities in magnetized cosmological spacetimes, either in the
Newtonian~\cite{10} or the GR limit~\cite{11}, ~\cite{12},
~\cite{13}, ~\cite{14}, ~\cite{15}, ~\cite{16}.

In the present article, we intend to re-construct the MHD
equations, both in the Newtonian and the GR limit and to use them
in order to study the finite-amplitude wave propagation in MHD
media. The reason to do so, actually relies on the thermal history
of the Universe.

According to the Standard Model~\cite{17}, ~\cite{18}, after nucleosynthesis
the Universe goes on expanding and cooling but nothing of great
interest takes place until $t \sim 10^{13}$ sec $(z \sim 10^3)$.
At that time, the temperature drops to the point where electrons
and nuclei can form stable atoms (recombination epoch).
During the so-called {\em radiation epoch}, photons
couple strongly with matter, the main constituent of which is in
the form of {\em plasma}. Interactions between the various
constituents of the Universal matter content along this period
include radiation-plasma coupling which is described by plasma
dynamics. On the other hand, the presence of plasma plays an
important role in shaping the radiation spectrum, something that
is fortified by the fact that there appear to exist cosmic
magnetic fields, of the order $10^{-6}$ G, which evidently have
not had enough time to evolve during the gravitating epoch of the
Universe~\cite{7}, ~\cite{12}. Therefore, although it is not
traditional to characterize the radiation epoch by the dominance
of plasma interactions, it may well be also called the {\em plasma
epoch}. It is the time-period during which the electromagnetic
interaction dominates among the four fundamental forces.

\section{MHD equations in an expanding Universe}

\subsection{The Newtonian MHD Theory}

During the plasma epoch, the MHD equations in the Newtonian theory
of gravity are given by~\cite{19}~\cite{20} \be \rh (\frac{D u_i}{d t}) =
-\frac{\partial{p}}{\partial{x_i}} + \rh
\frac{\partial{U}}{\partial{x_i}} + \frac{1}{4\pi}
\frac{\partial{(H_i H_j-\frac{1}{2}\delta_{ij}
H^2)}}{\partial{x_j}} \ee \be \frac{1}{\rh }\frac{D\rh}{dt} = -
\frac{\partial{u_i}}{\partial{x_i}} \ee \be \nabla^2 U = -4 \pi
G\rh \ee \be \frac{\partial{H_i}}{\partial t} =
\frac{\partial{(u_i H_j-u_j
H_i)}}{\partial{x_j}},~~\frac{\partial{H_i}}{\partial{x_i}}=0 \ee
In Eqs (1) - (4), Latin indices refer to the four-dimensional
spacetime (in accordance Greek indices refer to the
three-dimensional spatial section), $\dl_{ij}$ is the Kronecker
symbol and we have used the notation $$\frac{D}{dt} =
\frac{\partial }{\partial t} + u_i\frac{\partial}{\partial
{x_j}}$$ On the other hand, $\rh$ and $p$ denote the
energy-density and the pressure of a magnetized perfect fluid,
$H_i$ are the strength-components of the ambient magnetic field
and $U$ is the gravitational potential. Introducing
the scalars~\cite{21} \bea 2 \sg^2 & = & \sg_{ij}
\sg_{ij}~, ~~~2 \om^2 = \om_{ij} \om_{ij} \\
2 \tilde{\sg}^2 & = & \tilde{\sg}_{ij} \tilde{\sg_{ij}}~,~~~2
\tilde{\om}^2 = \tilde{\om}_{ij} \tilde{\om}_{ij} \eea where, \bea
&&\sg_{ij} = \frac{1}{2} (\frac{\partial{u_i}}{\partial {x_j}} +
\frac{\partial{u_j}}{\partial {x_i}} - \frac{2}{3} \dl_{ij}
\theta)~, ~~ \om_{ij} = \frac{1}{2} (\frac{\partial{u_i}}{\partial
{x_j}} - \frac{\partial{u_j}}{\partial {x_i}}) \nn \\
&&\tilde{\sg}_{ij} = \frac{1}{2}(\frac{\partial{H_i}}{\partial
{x_j}} + \frac{\partial{H_j}}{\partial {x_i}})~,~~
\tilde{\om}_{ij} = \frac{1}{2} (\frac{\partial{H_i}}{\partial
{x_j}} - \frac{\partial{H_j}}{\partial {x_i}}) \eea and $\theta =
\frac{\partial{u_i}}{\partial{x_i}}$ is the {\em expansion}
parameter, we obtain \be u_{i,j} = \sg_{ij} + \om_{ij} + {\theta
\over 3} \dl_{ij} \ee Combining Eqs (1) - (4) with Eq (8), we find
\bea &&\frac{1}{\rh}\frac{D^2 \rho}{dt^2}=\frac{\partial}{\partial
{x_i}} [\frac{1}{\rho} \frac{\partial{p^{*}}}{\partial{x_i}}] + 4
\pi G \rh \\
&+&\frac{H_j}{4 \pi \rho^2} \frac{\partial{H_i}}{\partial{x_j}}
\frac{\partial{\rho}} {\partial{x_i}} + 2 [\frac{2\theta^2}{3} +
(\sg^2- \frac{\tilde{\sg^2}}{4\pi\rh}) - (\om^2 -
\frac{\tilde{\om^2}}{4\pi\rh})] \nn \eea where, we have set $p^{*}
= p + \frac{H^2}{8\pi}$ to denote the {\em generalized pressure}
involved, due to the perfect fluid plus the magnetic field. Eq (9)
governs the finite-amplitude wave propagation in Newtonian MHD
theory. When $H_i = 0$, it reduces to the corresponding law
derived by Hunter~\cite{22}. On the other hand, using the notation
\be \frac{D H_i}{d t} = (\frac{\partial }{\partial t} + u_j
\frac{\partial }{\partial t}) H_i = H_j \frac{\partial }{\partial
x_j} - H_i \frac{\partial} {\partial x_j} \ee together with Eq
(8), we obtain \be \frac{D H_i}{dt} = (\sg_{ij} + \om_{ij} -
\frac{2}{3} \theta \dl_{ij})H_j \ee and multiplying Eq (11) by
$\mu H_i$, we end up an equation for the temporal evolution of the
magnetic field's energy density, namely \be \frac{D}{dt}
(\frac{\mu H^2}{8\pi}) = \frac{\mu}{4\pi}\sg_{ij} H_i H_j -
\frac{4\theta}{3} (\frac{\mu H^2}{8\pi}) \ee where, $\mu$ is the
permeability~\cite{11}.

{\em Summarizing}, the MHD dynamics in the Newtonian limit is
determined by the following system \be \rh(\frac{D u_i}{d t}) = -
\frac{\partial{p}}{\partial{x_i}} + \rh
\frac{\partial{U}}{\partial{x_i}} + \frac{1}{4\pi}
\frac{\partial{(H_i H_j - \frac{1}{2} \dl_{ij}
H^2)}}{\partial{x_j}} \ee \bea &&\frac{1}{\rh} \frac{D^2
\rho}{dt^2} = \frac{\partial}{\partial {x_i}} [\frac{1}{\rh}
\frac{\partial{p^{*}}}{\partial{x_i}}] + 4 \pi G \rh \\
&&+\frac{H_j}{4 \pi \rh^2} \frac{\partial{H_i}}{\partial{x_j}}
\frac{\partial{\rh}} {\partial{x_i}} + 2 [\frac{2\theta^2}{3} +
(\sg^2 - \frac{\tilde{\sg^2}}{4 \pi \rh}) - (\om^2 -
\frac{\tilde{\om^2}}{4 \pi \rh})] \nn \eea \be \nabla^2 U = -4 \pi
G\rh \ee \be \frac{D}{dt}(\frac{\mu H^2}{8\pi}) =
\frac{\mu}{4\pi}\sg_{ij} H_i H_j-\frac{4\theta}{3}(\frac{\mu
H^2}{8\pi}) \ee

\subsection{The Relativistic MHD Theory}

To derive the general relativistic MHD equations, we start up with
the Einstein field equations~\cite{19} \be {\cal R}_{ab} -
\frac{1}{2} g_{ab} {\cal R} = \kp T_{ab}~,~~~ \kp = - \frac{8\pi
G}{c^4} \ee where, ${\cal R}_{ab}$ is the Ricci tensor, ${\cal R}
= g_{ab} {\cal R}^{ab}$ is the corresponding scalar curvature and
$T_{ab}$ is the energy-momentum tensor, representing the average
state of the matter-energy content. Eq (17), together with the
Bianchi identities \be ({\cal R}^{ab} - \frac{1}{2} g^{ab} {\cal
R})_{;ab} = 0, \ee result in the conservation law $T_{\; \;
;b}^{ab}=0$, where, the semicolon denotes covariant derivative. In
what follows, we perform our calculations in the system of
geometrical units, where $G = c = 1$, admitting that the fluid's
four-velocity satisfies the condition $u_{a}u^{a} = - 1$; i.e. we
place ourselves in a coordinate system comoving with the fluid.

Again, we consider a magnetized perfect fluid source, determined
by the following energy-momentum tensor \be T^{ab} = (\rh +
\frac{H^2}{2}) u^a u^b + (p + \frac{H^2}{2}) h^{ab} - H^a H^b \ee
where, $$ h^{ab} = g^{ab} + u^a u^b $$ is the projection tensor
and $$ \eps = \rh + \rh \Pi $$ is the {\em total energy density},
due to the mass-energy content and the internal motions $(\rh
\Pi)$. By analogy to the corresponding evaluation in the Newtonian
limit, we introduce the {\em expansion} $\theta = u_{;a}^a$, the
{\em shear} $\sg_{ab} = h_a^ch_b^d u_{(c;d)} - \frac{1}{3} \theta
h_{ab}$ and the {\em twist} $\om_{ab} = h_a^c h_b^d u_{[c;d]}$
(round brackets denote symmetrization while square brackets
antisymmetrization), to write the covariant derivative of the
fluid's four-velocity in the form \be u_{a;b} = \sg_{ab} +
\om_{ab} + \frac{1}{3} \theta h_{ab} - \dot{u}_a u_b \ee where, we
have set $\dot{u}^a = u_{;c}^a u^c$.

Accordingly, the MHD dynamics in the GR limit is determined by the
following system~\cite{11} \bea &&x\dot{u}^a + \dot{x} u^a + x
\theta u^a + (p + \frac{H^2}{2})_{;b} g^{ab} = (H^a H^b)_{;b} \\
&&(\rh - \frac{H^2}{2})_{;ab} u^a u^b = h^{ab} (p +
\frac{H^2}{2})_{;ab} + 2 {d \over dt}(H^2 \theta) - \nn \\
&&-(H^a H^b)_{;ab} + 2 x (\frac{2\theta^2}{3} + \sg^2 -
\om^2-\dot{u}^a \dot{u}_{a}) + \nn \\
&&\frac{x}{2} (\rh + 3p + H^2) + 2 \dot{u}_{a}(H^aH^b)_{;b} +
(H^2)_{;a} \dot{u}^a \eea and \be {\frac{\mu {\dot{H}}^2}{8\pi}} =
\frac{\mu} {4\pi} \sg_{ij} H^i H^j - \frac{4 \theta}{3}(\frac{\mu
H^2}{8\pi}) \ee where, we have set $x = \rh + p + H^2$. Eqs (21) -
(23) may be applied to the investigation of perturbation effects
in curved spacetime and, hence, to the discussion of linearized
stability criteria. In order to obtain the perturbed expression of
the above equations, we follow the well-established method imposed
by Hawking~\cite{23}.

\section{Part I: MHD phenomena along the plasma epoch}

To discuss MHD phenomena in the early stages of the Universe, we
use a solution to the Einstein-Maxwell equations, first developed
by Thorne~\cite{24} and later discussed by Jacobs~\cite{25}, which
contains an ambient magnetic field. The model is homogeneous and
has two equivalent "transverse" directions and one inequivalent
"longitudinal" direction at each point of the spacetime. The magnetic
field is frozen into the matter content and is directed along the
longitudinal direction ($\hat{z}$-direction). We consider that the
matter-content filling this model is in the form of the perfect
fluid (19), which obeys the equation of state $p = \gm \rh$, with
$1/3 \leq \gm \leq 1$. In this case, the line-element
reads~\cite{24} \be ds^2 = - dt^2 + t (dx^2 + dy^2) + t^m dz^2 \ee
where, $m = 2 \frac{(1 - \gamma)}{(1 + \gamma)}$. With the aid of
Eq (24), we may apply the GR limit of MHD dynamics to discuss
stability of Bianchi-Type I cosmological models. In this case, we
depart from the traditional approach, adopting an anisotropic
cosmological model, in which the inherent magnetic field is
responsible for the initial anisotropy of the Universe.

\subsection{Excitation of low-frequency plasma waves}

To deal with small-amplitude waves, we first assume a background
situation, representing a uniform plasma in curved spacetime, in
the presence of an ambient magnetic field. In our case, this
situation is described by the solution (24), together with the
energy-momentum tensor (19). Accordingly, we introduce first-order
perturbations to the MHD equations (21) - (23) and taking into
account the so-called Cowling approach~\cite{26}, which admits
that $\dl g_{ab}=0$, we neglect all terms higher or equal than the
second order. Eventually, we search for solutions to the
linearized MHD equations in which all the perturbation quantities
are written in the form of a plane-wave, with angular frequency
$n$ and a wave-vector $\vec{k}=(0,0,k^3)$ parallel to the magnetic
field, $\vec{H}=(0,0,H^3)$. Accordingly, \be \dl \eps, \dl p, \dl
H^a, \dl u^a \sim \exp{i(nt-kz)}. \ee

Upon consideration of the perturbed MHD equations~\cite{23}, after
straightforward but tedious calculations, we obtain the equation
governing the density fluctuations~\cite{11} \bea
&&\frac{\partial^2 \dl \rh}{\partial t^2} + A_1 \frac{\partial \dl
\rh}{\partial t} - \dl p_{;ab} h^{ab} - A_2 c_s^2 \frac{\partial^2
\dl \rh}{\partial z^2} - [A_3 + c_s^2 A_4] \dl \rh \nn \\
&&= c_1 \dl u^0 + c_3 \dl H_{;3}^3 + F_1 \dl H^3 + F_2 \dl
H_{;ab}^3 + F_3 \dl H_{;b3}^3 \eea where, $c_s^2 = \dl p / \dl
\rh$ is the speed of sound and $A_i, c_i, F_i$ $(i=1,2,3)$, are
all functions of $t$ and $\gm$, the explicit form of which is given
in~\cite{11}. Inserting Eqs (25) into Eq (26), we obtain a secular
equation of the form \be n^2-i n A - S_1 - i k S_2 = 0 \ee where,
$A$, $S_1$ and $S_2$ are complicated expressions of the background
energy density and the pressure, as well as of the shear and the
expansion of the Universe, together with the magnetic field's
strength (see Appendix A). Eq (27) results in \be n_1 = \ta
\cos{\om/2} + i [\frac{A}{2} + \ta \sin{\om/2}] \ee and \be n_2 =
- \ta \cos{\om/2} +i [\frac{A}{2} - \ta\sin{\om/2}] \ee where, we
have set \be \ta = 4 \sqrt{(S_1 - \frac{A^2}{4})^2 + k^2 S_2^2}
\ee and \be \cos{\frac{\om}{2}} = \frac{kS_2}{\ta}~,
~~~\sin{\frac{\om}{2}} = \frac{4 S_1 - A^2}{4\ta} \ee According to
Eqs (28) and (29), the medium remains stable when the frequency
equals to $n_1$. In this case, the magnetic field's perturbation,
$\dl H^3$, turns out to be an {\em Alfven wave}. On the other
hand, for $n = n_2$, the medium becomes unstable and $\dl H^3$
grows in time. We identify this instability with the well known
{\em Parker's instability}~\cite{27}.

There are also other potential instabilities that can be traced
within the context of general relativistic MHD. In particular, a
spectrum of magnetized sound waves may be excited and form
large-scale damped oscillations in the expanding Universe (e.g.
see~\cite{16}). The characteristic frequency of the excited waves
is slightly shifted away from the sound frequency and the shift
depends on the strength of the primordial magnetic field. This
magnetic-field-dependent shift, may have an effect on the acoustic
peaks of the CMRB.

One of the main scopes of this study, is to examine the spectrum
of the unstable low frequency plasma waves in the anisotropic
cosmological model (24). Under the assumption that the magnetic
field lays along the $\hat{z}$-direction and the perturbations are
plane-waves propagating also in the $\hat {z}$-direction, the
perturbed general relativistic MHD equations, result in the {\em
dispersion relation} \bea &&-n^2 + k^2 c_s^2 - (1 + c_s^2)
[-\frac{8}{3} \theta^2 + 2 \sg^2 + \frac{1}{2} (\rh + 3p) \nn \\
&+&\frac{2 \theta H_{,0}^2}{\rh + p} + \frac{4 \theta^2 H^2}{3(\rh
+ p)} - \frac{2H^2}{3}] - \frac{1}{2}(p + \rh)(1 + 3 c_s^2) \nn \\
&-& i n [4 \theta + \frac{2H_{,0}^2}{(\rh + p)}- \frac{4\theta
H^2}{3 (\rh + p)}] \nn \\
&=&(\mathcal{R}_1 + i n \mathcal{R}_2) \frac{\dl
H^3}{\mathcal{\dl} \eps} \eea where, we have set \be \frac{\dl
H^3}{\dl \eps} = \frac{H^3}{4 \pi \mathcal{D} (2 - \gm)} [1 -
\Lm_1 - i n \Lm_2], \ee and $\mathcal{R}_1$, $\mathcal{R}_2$,
$\Lm_1$, $\Lm_2$ and $\mathcal{D}$, are functions of the magnetic
strength $H$, the shear $\sg$ and the expansion $\theta$ of the
background solution (see Appendix B). At the early stages of the
evolution ($t \rarrow 0$), Eq (32) reads \be -n^2 [1 -
\frac{u_A^2}{2 - \gm}] + k^2 c_s^2 + \frac{k^2 u_A^2}{2 - \gm} +
T_1(t, \gm) + i n T_2(t, \gm) = 0 \ee where, $u_A$ is the Alfven
velocity and $T_1 (\gm , t)$, $T_2 (\gm , t)$ are complicated
expressions of $\gm$ and $t$, the exact form of which is given in
Appendix B. We re-write Eq (34) in the form $D_r+iD_i=0$ and,
assuming that $n = n_r + i n_i$, where, the real part of the
excited frequency is much larger than the corresponding imaginary
part ($n_r>>n_i$), we obtain:

\begin{itemize}

\item The real part of the excited frequency from the equation
$D_r=0$ \be n_{r}^2 [1 - \frac{u_A^2}{2-\gm}] = k^2 c_s^2 +
\frac{k^2u_{A}^2} {2 - \gm} + \frac{\tilde{T}_1 (\gm)}{t^2} \ee

\item The imaginary part of the excited frequency from the
relation~\cite{28} \be n_i = - {D_{i} (n_r , k) \over {\partial
D_{r}(n_r , k) \over
\partial n} \vert_{n = n_r}} \ee which yields
\bea n_i & = & \frac{T_2(\gm, t)}{2 (1 - u_A^2(t) / (2 - \gm))}
\nn \\
& = & \left(\frac{C(\gm)}{t} \right) \left(\frac{1}{2(1 - u_A^2(t)
/ (2 - \gm))}\right) \eea For $t \rarrow 0$, $n_i$ becomes
negative, decaying rapidly at late times. Therefore, we verify
that the role of magnetic field is to make the expanding Universe
even more stable against the expected {\em Jeans-type
instabilities}.

\end{itemize}

On the other hand, the real part of the frequency is shifted away
from the frequency of sound \be n_r^2 = \frac{k^2 c_s^2 +
\frac{k^2 u_A^2}{2 - \gm} + \frac{\tilde{T}_1}{t^2}} {1 -
\frac{u_A^2(t)}{2 - \gm}} \ee In fact, as $\gm \rarrow 1/3$, the
magnetic field vanishes and the anisotropic model (24) approaches
the weakly-magnetized Friedmann - Robertson - Walker (FRW) model,
which is used extensively in the literature (e.g. see ~\cite{31}
and references therein). In addition, the characteristic frequency
$n_r$ approaches the value $n_r = k c_s + \Dl n_r (u_A)$. The
presence of the Alfven velocity in Eq (38), may be responsible for
the distortion of acoustic peaks, as it was pointed out first
by~\cite{12}. Therefore, a spectrum of low frequencies will be
excited with frequency $n_r$.

\subsection{Evolution of the density fluctuations}

Finally, we examine the temporal evolution of the density
fluctuations. A linear density perturbation has the form \be \dl
\rh \sim \left[ \rh_0 e^{|B(\gm, t)|} \right] e^{i (n_r t - k z)}
\ee where, $$B(\gm, t) = {C(\gm) \over 2 {1 - u_A^2 (t) \over 2 -
\gm} }$$ The function $|B(\gamma,t)|$ decays as time grows or,
equivalently, as $\gm$ decreases. In particular, as $t \rarrow 0$,
which corresponds to $\gm \rarrow 1$, we have $u_A \rarrow 1$ and
$|B(\gm, t)| \rarrow \infty$. On the other hand, for $t \rarrow
\infty$, which results in $\gm \rarrow 1/3$, we obtain $u_A
\rarrow 0$ and $|B(1/3,t)| \rarrow (3/2)$. These results indicate
that, at early times $(t \rarrow 0)$, the perturbations' amplitude
becomes large and the excited waves will form a {\em spectrum of
damped oscillations} within the magnetized cosmological model. We
may search for the source of this damping by analyzing the
functional form of $T_2(\gm, t)$ in Eq (34). In the absence of
magnetic fields ($\gm = 1/3$), we obtain $T_2 = - 4 \theta$ and
the expanding plasma stabilizes the ion-acoustic waves.

On the other hand, if the cosmological model is {\em static} $(A =
const.$ and $\theta \sim {\dot{A} \over A} = 0)$, stabilization of
the magnetosonic waves can be caused by the magnetic pressure. In
a magnetized expanding Universe [Eq (24)] both processes will be
combined to damp the large-amplitude oscillations. Assuming, for
the sake of agreement, that at a some time $t = const.$ the
Universe becomes static, a {\em Jeans-like instability} sets in
(see also ~\cite{29}). Therefore, we realize that the damping is
caused by the {\em free energy}, available both in the plasma and
the magnetic field, during the early stages of the Universal
expansion.

{\em Summarizing}, we have analyzed the small-amplitude wave
propagation in the early Universe, using an anisotropic
cosmological model with an {\em inherent} magnetic field towards
the $\hat{z}-$direction, which is frozen into a perfect fluid with
equation of state $p = \gm \rh$ (where, $1/3 \leq \gm \leq 1$).
For $\gm \rarrow 1/3$, the anisotropic model used, approaches the
weakly magnetized FRW model, in which the magnetic field is a
small linear perturbation (e.g. see~\cite{14}, ~\cite{30}).

\section{Part II: MHD phenomena of cosmological significance}

Newtonian Cosmology can be a quite good approximation for gravity,
at scales much smaller than the Hubble length. For this reason, it
is often used to describe our Universal neighborhood $(0.1 \leq z
\leq 1.5)$, within the context of Observational Cosmlogy; in
particular, as regards the local observations of cosmological
significance~\cite{18}. Within the context of Newtonian Cosmology,
we study the linear MHD behavior of a viscous magnetized fluid
with finite {\em conductivity} and generalize the {\em Jeans
instability criterion}. In particular, we look into the linear
evolution of small inhomogeneities in the cosmic medium and
examine how the magnetic field and the fluid's viscosity may
affect the characteristic scales of the gravitational
instabilities. Accordingly, we discuss the electrodynamic
properties of the collapsing fluid, the resulting amplification of
the magnetic field and the formation of unstable {\em
current-sheets}.

A central point to our discussion is the concept of {\em anomalous
resistivity}, which is triggered by electrostatic instabilities in
the plasma and may reduce substantially the corresponding
electrical conductivity. We argue that these changes in the
resistivity of the protogalactic medium might have lead to the
formation of strong electric fields during galactic collapse.
These fields can accelerate the abundant free electrons and ions,
to produce Ultra High Energy Cosmic Rays (UHECRs) during the
formation of protogalactic structures.

In Newtonian Cosmology, one deals with the following set of MHD
equations~\cite{19} \be \frac{\partial \rh}{\partial t} = - 3
\frac{\dot{a}}{a} \rh - \frac{1}{a} \vec{\nabla} \cdot (\rh
\vec{u}) \ee \bea \frac{\partial\vec{u}}{\partial t} & = & -
\frac{\dot{a}}{a} \vec{u} - \frac{1}{a} (\vec{u} \cdot
\vec{\nabla}) \cdot \vec{u} -
\frac{c_{\rm s}^2}{a \rh} \vec{\nabla} \rh + \frac{1}{a} \vec{\nabla} \ph \nn \\
& + & \frac{1}{4 \pi a \rh} (\vec{\nabla} \times \vec{B}) \times
\vec{B} + \frac{\nu}{a^2 \rh} \nabla^2 \vec{u} \eea \be \nabla^2
\ph = - 4 \pi G a^2 \rh \ee \be \frac{\partial \vec{B}}{\partial
t} = - 2 \frac{\dot{a}}{a} \vec{B} + \frac{1}{a} \vec{\nabla}
\times (\vec{u} \times \vec{B}) + \frac{\et}{a^2} \nabla^2 \vec{B}
\ee \be \vec{\nabla} \cdot \vec{B} = 0 \ee where, $a(t)$ is the
{\em radius} of the Newtonian Universe and $\ph$, stands for the
gravitational potential.

\subsection{Magnetically induced anisotropies in the gravitational collapse}

Once again, we perturb Eqs (40) - (44) around a zeroth order
background solution, representing a uniform plasma in the presence
of an ambient magnetic field
$$\rh = \rh_0 + \rh_1~,~~~ \vec{B} = \vec{B}_0 + \vec{B}_1~,~~~\ph
= \ph_0 + \ph_1~,~~~ \vec{u} \neq 0$$ We assume plane-wave
perturbations and take the time-derivative of Eq (41). The real
part yields \bea \ddot{\vec{u}} & = & - \left ( H + \frac{\nu
k^2}{a^2 \rh_0} \right )
\dot{\vec{u}} \nn \\
&+& \left [ 8 \pi G \rh_0 - \frac{k^2}{a^2} ( c_{\rm s}^2 + u_A^2
+ \frac{\nu H}{\rh_0} ) \right ] \vec{u} \eea Furthermore, we
assume that the perturbations propagate {\bf (a)} parallel to the
magnetic field and {\bf (b)} perpendicular to the magnetic field.
Accordingly, we end up with two {\em critical wavelengths},
associated to the corresponding Jeans scales, namely \be
\lm_{\perp} \simeq \sqrt{\frac{c_{\rm s}^2 + u_A^2 + \nu H /
\rh_0}{8 \pi G \rh_0}}~, ~~~ \lm_{\parallel} \simeq
\sqrt{\frac{c_{\rm s}^2 + \nu H / \rh_0}{8 \pi G \rh_0}} \ee
where, $H = \frac{\dot{a}}{a}$ is the Hubble parameter, $\nu$ is
the viscosity of the plasma fluid and $\et$ is the corresponding
resistivity. From Eq (46), it is evident that the magnetic field
{\em induces} a degree of anisotropy in the gravitational collapse
and, therefore, one expects gradual formation of turbulent motions
within the magnetized medium. These motions, in turn, contribute
additional terms to the viscosity of the fluid. In the
post-recombination epoch, the viscosity due to turbulent motions
reads \be \nu_{turb} \sim \rh_0 u_1 l_{mix}\ee where, $l_{mix}$
the turbulent mixing length. Assuming that
\begin{itemize}

\item $u_1$ reaches values close to $u_A$

\item $l_{mix} \ll \lm_{\perp}$ and

\item $T_{rec}$ is known

\end{itemize}
we end up with the condition \be c_{s}^2 < \frac{\nu H}{\rh_0} <
u_A^2 \ee At the present epoch, we adopt the typical
values~\cite{18}: For the magnetic field, $B \leq 10^{-7}$ G; for
the energy-density, $\rh_0 \geq 10^{-29}$ ${\rm gr} \cdot {\rm
cm}^{-3}$ and for the Hubble parameter, $H = 100 \, h$ ${\rm
km}\cdot{\rm sec}^{-1}\cdot{\rm Mpc}^{-1}$, where $0.4 \leq h \leq
1$, we find \be {\lm_{\perp} \over \lm_{\parallel}} \sim
\sqrt{{c_{\rm a}^2 \rh_0 \over \nu_{turb} H}} \gg 1 \ee with
$\lm_{\perp}$ being of the order of a (comoving) Mpc. Following
the standard structure formation scenarios, the initial collapse
of a large structure will be followed by successive fragmentation
into smaller scale formations with characteristic lengths $\lm \ll
\lm_{\perp}$. Moreover, as we will outline next, the magnetic
field which is trapped into the gravitating medium, will be
increased even further, due to the anisotropic collapse. In the
case of an almost spherically symmetric collapse, linear
inhomogeneities in the magnetic field's energy density amplify in
tune with those in the density of the matter, so that
$$\dl B^2 \propto \dl \rh$$ where, $\dl B^2 = B_1^2/B_0^2$ and
$\dl \rh = \rh_1/\rh_0$ Therefore, even in the spherically
symmetric case, the formation of matter condensations in the
post-recombination Universe signals the amplification of any
magnetic field that happens to be present at that time~\cite{32}.

We have already seen that the anisotropic magnetic field induces
some degree of anisotropy in the gravitational collapse of the protogalactic
structure. On the other hand, this magnetically induced anisotropy
will react back to affect the evolution of the field itself. The
evolution of the magnetic field during the nonlinear regime of a
generic, non-spherical protogalactic collapse, has been considered
by a number of authors (e.g. see ~\cite{31} and references there
in). The approach is both analytic and numerical and agree that
{\em shear effects} increase the final strength of the magnetic
field, while confining it onto the protogalactic plane. Compared
to the magnetic strength of the spherical collapse scenario, this
anisotropic increase is larger by, at least, one order of
magnitude and, therefore, protogalactic structures can be endowed
with magnetic fields considerably stronger than those previously anticipated.

So far, we have seen how the magnetic presence modifies the way
the gravitational collapse may proceed, by changing the overall
stability of the magnetized fluid. This, in turn, affects the
evolution of the magnetic field itself and may trigger a chain of
nonlinear effects on certain scales.

At next, we will see that, selective amplification of certain
perturbation modes, plays an important role during the nonlinear
stages of protogalactic collapse, helping instability to reach its
{\em saturation point}.

\subsection{Amplification of the induced electric currents}

In the presence of magnetic fields there is an electric current
induced by $\vec{B}$, namely \be \vec{J} = \frac{c}{4 \pi}
\vec{\nabla} \times \vec{B} = \frac{c}{4 \pi} \al \vec{B} \ee
where, $\vec{\nabla} \times \vec{B} =\al \vec{B}$ and $\al$
measures the magnetic torsion (see~\cite{33} for details). One
expects that, initially, $\al$ is small. However, the subsequent
fragmentation of the protogalactic cloud to scales with  $\lm \ll
\lm_\perp$ will increase $\vec{\nabla} \times \vec{B}$ and
strengthen the induced current. In particular, for $B \sim
10^{-7}$ G, we obtain $J \sim 10^2 \al$ which may reach
appreciable strengths for reasonable values of $\alpha$.

When the induced current exceeds a critical value, namely $J_c
\sim \rh (e/m_p)c_{\rm s}$, where, $c_{\rm s} \sim 10^4
\sqrt{T(^\circ K)}$ cm/sec, $m_p$ is the ion mass and $e$ is the
electron charge, the excitation of low frequency electrostatic
turbulence will increase the resistivity of the medium by several
orders of magnitude~\cite{34}. In fact, even small values of $\al$
may lead to $J > J_c$, thus making the plasma electrostatically
unstable. This effect, which is known as {\em anomalous
resistivity}, can be explained by the development of current-driven
electrostatic instabilities in plasma, that could lead to
the excitation of a variety of waves and oscillations. Absorption
of these waves by the ions, results in a transfer of momentum from
the electrons to the ions, along with the usual momentum loss from
the former species to the latter. The average momentum loss by the
electron per unit time can be written as an effective collision
term, in the form \be n m_e \nu_{eff} \vec{u}_{e} = - \vec{F}_{fr}
\ee where, $F_{fr}$ is the average {\em friction force} and $n =
\rh/m_p$ is the ambient number density of the plasma particles.
The friction force is proportional to both the linear growth rate
of the electrostatic waves ($\gm_k$) and the energy of the excited
waves ($W_k$). The effective collision frequency is estimated to
be $\nu_{eff} = \om_e (W_{sat} / k_B T)$. Then, the anomalous
resistivity will be \be \et_{an} \sim \frac{\nu_{eff} c^2}{4 \pi
\om_{e}^2} \sim \left(\frac{W_{sat}}{k_B T} \right) \left(
\frac{c^2}{4 \pi} \right) \frac{1}{\om_e} \ee where, $\om_e = 5.6
\times 10^4 \sqrt{n}~{\rm sec}^{-1}$ is the plasma frequency,
$k_B$ is the Boltzman constant and $(W_{sat}/k_B T) \sim
10^{-6}-10^{-4}$ is the saturated level of the electrostatic waves
~\cite{34}. For certain types of current-driven waves, the
anomalous resistivity is several orders of magnitude above the
classical one, as confirmed in numerous laboratory
experiments~\cite{35}.

This sudden switch to high electrical resistivity leads to the
formation of strong electric currents and therefore to a fast
magnetic dissipation and intense plasma heating. The electric
fields induced by the gravitational collapse will be \be E \sim
{u_A \over c} B + \et_{an} J_c \sim \et_{an} J_c \ee provided that
${u_A \over c} \ll 1$. According to this scenario, the
gravitational collapse of a magnetized dust cloud amplifies the
magnetic field and (indirectly) generates strong electric currents,
which form large-scale current-sheets localized on the
protogalactic plane.

The anisotropy of the collapse enhances the local currents even further
and, eventually, drives the resistivity towards anomalously high
values. The inevitable result is strong electric fields
accelerating the abundant free electrons. In particular, the
energy gained by an electron travelling a length, $\lm \sim
10^{-3} \lm_{\perp} \ll \lm_{\perp}$ is $W_{kin} \sim e E \lm$ and
the relativistic factor, $\gm = [1-(v/c)^2]^{-1}$ is given by \bea
\gm & = & \frac{W_{kin}}{m_e c^2} \sim \frac{e E \lm_\perp} {10^3
m_e c^2} \sim \frac{10 e^2 \et_{an} \sqrt{T}B}{m_e m_p c^2
G^{1/2}} \Rightarrow \nn \\
\gm & \sim & 10^{14} \left ( \frac{n}{10^{-4} {\rm cm}^{-3}}
\right )^{-{\textstyle{1\over2}}} \left ( \frac{T}{1^{\circ}K}
\right )^{\textstyle{1\over2}} \left ( \frac{B}{10^{-7}{\rm G}}
\right ) \eea in CGS units. Taking into account that $$\lm_\perp
\sim B/n m_p \sqrt{G},$$ $$c_{\rm s}^2 < \nu H/ \rh <u_A^2$$ and,
through Eq (52), $$W_{sat} /k_B T \sim 10^{-6},$$we find that the
typical energy gain by a free electron may reach extremely high
values within short timescales ($t_{acc} \sim \lm / c \ll
\lm_\perp / c \sim 10^6$~yrs), even for relatively weak magnetic
fields. It is straightforward to extend this process to proton
acceleration and show that protogalactic collapse may be the
source of UHECRs.

We also anticipate a few particles travelling across several
fragments adding up to a scale comparable to the Jeans length.
Many particles will drift in and out the primordial
current-sheet, as well as the associated strong electric fields.
Through synchrotron emission, particles gain and lose energy
continuously. Provided that fragmentation has already taken place,
these particles may diffuse along the different current-sheets,
forming the observed power-law distribution. The role of
unstable current-sheets on the acceleration of cosmic rays along
the giant radio-galaxies, has already been pointed out in the
literature~\cite{36}.

{\it Summarizing}, the suggestion made here is that, in the
context of Newtonian Cosmology, production of UHECRs may have
started almost simultaneously with the formation of galaxies,
through the electrodynamic characteristics of the gravitational
instability. This process may be continued until today, since the
previously described instability is active on all cosmic scales.

\newpage

\section{Conclusions}

{\bf Part I:}

\begin{itemize}

\item[1] The Jeans instability, established in an Einstein
Universe with a very particular cosmological constant~\cite{29},
is absent in the presence of magnetic fields, which act as
stabilizing forces.

\item[2] An anisotropic magnetized Universe re-enforces the
stability, initially found in the FRW model, against Jeans-like
instabilities.

\item[3] A spectrum of low frequency and large amplitude damped
oscillations may appear in the early Universe, with a
characteristic frequency $n_r = k c_s + \Delta n$, where, $\Delta
n$ depends on the strength of the primordial magnetic field.

\end{itemize}

Because of the above results, we come to the following
conclusions:

\begin{itemize}

\item The magnetic-field-dependent shift $(\Delta n)$, may cause
measurable distortion on the acoustic peaks of the CMRB (see also
~\cite{12}, ~\cite{15}, ~\cite{30}).

\item The amplitude of the excited waves is substantially large in
the early stages of the Universal evolution and gradually decays.

\item If the Universe has passed through a strongly magnetized
anisotropic phase, the excited waves may be responsible for the
formation of large-scale fluctuations, as it has been shown in 2-D
numerical simulations~\cite{37}.

\end{itemize}

{\bf Part II:}

\begin{itemize}

\item[1] We have found that, a magnetic field frozen onto a fluid
of finite viscosity, may alter the standard picture of the
gravitational instability.

\item[2] We have discussed how the preferential, anisotropic
magnetic amplification may increase the induced electric currents
on a plane perpendicular to the main axis of the collapse.

\item[3] These gravitationally induced current-sheets will, in
turn, trigger electrostatic instabilities, which may lead to
anomalously high values of the resistivity and, consequently, to
strong electric fields.

\item[4] These electric fields can be strong enough, in order to
accelerate free electrons up to ultra high energies, producing
UHECRs.

\end{itemize}

To the best of our knowledge, this is the first time that a direct
connection between gravitational instability and cosmic-ray
acceleration has been suggested and discussed.

\vspace{.5cm}

The authors would like to thank Dr C G Tsagas for helpful
discussions. Financial support from the Greek Ministry of
Education under the Pythagoras programm, is gratefully
acknowledged.

\section*{Appendix A}

The expressions $A$, $S_1$ and $S_2$, appearing in Eq (27), are
written in the form: \begin{eqnarray*} &&A = \frac{1}{\rh + p}
[\frac{13}{3} H^2
\theta - (\rh + p) \theta \nn \\
&+&\frac{35}{3} H_{,z} H_{,0}^z + \frac{1}{2} g^{zz} g_{zz,0} H^2
\frac{2 \rh + 2 p +H^2}{\rh + p + H^2}] \end{eqnarray*}
\begin{eqnarray*}
&&S_1 = -\frac{\rh}{2} (1 + c_s^2) (1 + 3 c_s^2) \nn \\
& + & k^2 [c_s^2 + \frac{H^2 - (\rh + p + 3H^2) H_z}{(\rh + p + H^2)}] \nn \\
& - & c_s^2 H^2 - \frac{1}{2} [\rh + 3 p + \frac{4 \theta^2}{3} + 4 \sg^2 +
\frac{4}{3} H^2 - \frac{8}{3} H^z H_{z;0} \nn \\
& + & \frac{2 \theta}{\rh + p}(H_z H_{;0}^z + \frac{13}{3} \theta H^2 + \frac{32}{3}
\theta H_z H_{;0}^z) \nn \\
& + & \frac{1}{2} g^{zz} g_{zz;0} H_z (2 \rh + 2 p + H^2) (\rh + p + H^2)^{-1}]
(1 + c_s^2) \nn \\
& - & \{ H_{;00}^z - 6 \theta H_{;0}^z - H^z [3 H^2 + 6 \sg^2 - 6 \theta^2 +
\frac{\rh - 5 p}{3} \nn \\
& + & \frac{1}{2} g^{zz} g_{zz,0} (\frac{1}{2} g^{zz} g_{zz,0} - \theta)
\frac{2 \rh + 2 p + H^2}{\rh + p + H^2}] \nn \\
& - & \frac{8}{3} H^2 H_{;0}^z \} - \{2 H_{;0}^z H_{z;0} + H_z H_{;00}^z + \theta
H_z H_{;0}^z \nn \\
& + & 2H^2 [2 \sg^2 + \frac{\theta^2}{3} + \frac{1}{2}(\rh + p + H^2)] \nn \\
& + & \frac{\rh_{,0}}{\rh + p} (H_z H_{;0}^z + \frac{13}{3} \theta H^2 +
\frac{32}{3} \theta H_z H_{;0}^z) \nn \\
& + & \frac{2 H_z H_{;0}^z}{\rho + p + H^2} (p + \frac{H^2}{2})_{;0} - \theta
\rho_{;0} \nn \\
& + & g^{zz} g_{zz,0} [5 H^2 \theta - H_z H_{;0}^z - H^2 u_{;z}^z
+ \frac{H^2}{\rh + p + H^2} (p + \frac{H^2}{2})_{;0} \nn \\
& - & H^z \frac{2 \rh + 2 p + H^2}{\rh + p + H^2}(\frac{H^z
\rh_{,0}}{\rh + p} -H_{,0}^z)]\} \end{eqnarray*} \begin{eqnarray*}
&&S_2 = H_z H_{;0}^z + H^2u_{;z}^z - 5 H^2 \theta -
\frac{H^2}{\rh + p + H^2} (p + \frac{H^2}{2})_{,0} \nn \\
& + & g^{zz} H^2 \frac{2 \rh + 2 p + H^2}{\rh + p + H^2}
\end{eqnarray*}

\section*{Appendix B}

The functions $\mathcal{R}_1$, $\mathcal{R}_2$, $\Lm_1$, $\Lm_2$ and
$\mathcal{D}$, appearing in Eq (32) are:\\
\begin{eqnarray*} \mathcal{R}_1 & = & [H_{3,00} - 4 (\theta H_{3,0} +
\theta_{,0} H_3) - 2H_{,0}^3 \Gamma_{33}^0] \\
& + & H_3 [ - \frac{4}{3} H^2 + \frac{4}{3} \theta^2 + 2 \sg^2 + 2
(\rh + 2 p) ] \\
& + & g^{33} [ - 2 \Gamma_{33,0}^0 - 2 \Gamma_{33}^0 \Gamma_{b0}^b
+ 2 \Gamma_{33}^0 \Gamma_{30}^3 ] H_3 \\
& - & (n^2 + k^2) H_3 \end{eqnarray*}
\[\mathcal{R}_2=2H_{3,0}-4\theta H_3-\Gamma_{33}^0 H_3 g^{33}\]
\begin{eqnarray*} \Lm_1 & = & \frac{1}{\mathcal{D}} [ k^2 (H^3)^2 +
(2 \gm - 1) \theta [2 x \Gamma_{30}^3 + (p + H^2 / 2 )_{,0}] \\
& - & k^2 c_s^2 g^{33} (\eps + p)] \end{eqnarray*}
\[\Lm_2 = (2 \gm - 1) x \theta,\]
\begin{eqnarray*}
\mathcal{D} & = & - n^2 x + k^2 (H^3)^2 - (2 - \gm) \theta \left [
2 x \Gamma_{30}^3 + (p + \frac{H^2}{2})_{,0} \right] \\
& - &i n \left [ 2 x \Gamma_{30}^3 + p + \frac{H^2}{2} - (2 - \gm)
x \theta \right] \end{eqnarray*}

Finally, the functions $T_1$ and $T_2$, appearing in Eq (34), have
the following functional form:
\[ T_1(\gm , t) = \frac{\tilde{T}_1(\gm)}{t^2}\]
\begin{eqnarray*}\tilde{T}_1(\gm) & = & - (9 \gm^6-993 \gm^5 + 4842 \gm^4
-6818 \gm^3 +2889 \gm^2 \\
& - & 461 \gm + 788) / (6 (\gm - 2) (\gm - 3)^2 (1 + \gm)^2)
\end{eqnarray*} and \begin{eqnarray*}
T_2 (\gm, t) & = & - 4 \theta + \frac{4 \theta}{3} \left (
\frac{H^2}{4 \pi(\rh + p)} \right) + 4 \theta \left ( \frac{H^2}{4
\pi (2 - \gm)} \right) \\
& - & \frac{2H_{,0}^2}{4 \pi(\rh + p)} - \frac{2H_{3,0} H_3}{4 \pi
\rh (2 - \gm)} \\
& + & \frac{(1 - \gm)}{(\gm + 1) t} \left ( \frac{H^2}{4 \pi \rh
(2 - \gm)} \right ) \end{eqnarray*} where, $\theta = - 2 /
[(1+\gamma)t]$. The function $T_2$ depends on $\gm$ and t, since
the magnetic field, the energy density and the pressure are all
functions of $\gm$ and t. In other words, $T_2 (\gm, t)$ has the
form
\[ T_2(\gm,t) = \frac{C(\gm)}{t},\] where,
\[C(\gm) =
\frac{45 \gm^4 - 36 \gm^3 - 126 \gm^2 + 212 \gm -127} {3 (\gm - 3)
(\gm - 2) (\gm + 1)^2}\] We can easily show that $T_1(\gm, t)$ is
positive and $T_2(\gm, t)$ is negative, for all the allowed values of
$1/3 \leq \gm \leq 1$.


\newpage

\begin{thebibliography}{40}

\bibitem{1} Punsly B, 2001, {\it Black hole
Gravitohydromagnetics}, Springer, Berlin
\bibitem{2} Jackson J D, 1975, {\it Classical Electrodynamics},
Wiley, New York
\bibitem{3} Carot J and Tupper B O J, 1999, Phys Rev {\bf D 59},
124017
\bibitem{4} Dimmelmeier H, Font A J and M\"uller E, 2002, A \& A
{\bf 393}, 523
\bibitem{5} Fryer C L, Holz D E and Hughes S T, 2002, ApJ {\bf
565}, 430
\bibitem{6} Baumgarte T W and Shapiro S L, 2003, Ap J {\bf 585},
930
\bibitem{7} Kronberg P P, 1994, Rep Prog Phys {\bf 57}, 325
\bibitem{8} Kronberg P P, Perry J J and Zukowski E L H, 1992, ApJ
{\bf 387}, 528
\bibitem{9} Wolfe A M, Lanzetta K M and Oren A L, 1992, ApJ {\bf
388}, 17
\bibitem{10} Hacyan S, 1983, ApJ {\bf 273}, 421
\bibitem{11} Papadopoulos D B and Esposito F P, 1982, ApJ {\bf 257}, 10
\bibitem{12} Adams J, Danielsson U H, Grasso D and Rubinstein H, 1996,
Phys Lett {\bf B 388}, 253
\bibitem{13} Barrow J D, Ferreira P G and Silk J, 1997, Phys Rev {\bf
D 78}, 3610
\bibitem{14} Durrer R, Kahniashvil T and Yates A, 1998, Phys Rev {\bf
D 58}, 123004
\bibitem{15} Jedamzik K, Katalinic V and Olinto A, 2000, Phys Rev
Lett {\bf 85}, 700
\bibitem{16} Papadopoulos D B, Vlahos L and Esposito F P, 2002, A \& A,
{\bf 382}, 1
\bibitem{17} Weinberg S, 1972, {\it Gravitation and Cosmology}, Wiley,
New York
\bibitem{18} Narlikar J V, 1983, {\it Introduction to Cosmology}, Jones
and Bartlett Publishers Inc, Boston
\bibitem{19} Zel'dovich Ya B and Novikov I D, 1983, {\it The Structure
and the Evolution of the Universe}, Univ of Chicago Press Ltd,
London
\bibitem{20} Chandrasekhar S, 1961, {\it Hydrodynamic and Hydromagnetic
Stability} Oxford Univ Press, Oxford
\bibitem{21} Ellis G F R, 1971, {\it Lectures in General Relativity and
Cosmology}, in {\em Proceedings of the International School of
Physics Enrico Fermi XLVII}, R K Sachs (Ed)
\bibitem{22} Hunter C, 1964, ApJ {\bf 139}, 570
\bibitem{23} Hawking S W, 1966, ApJ {\bf 145}, 544
\bibitem{24} Thorne K S, 1967, ApJ {\bf 148}, 51
\bibitem{25} Jacobs K C, 1969, ApJ {\bf 155}, 379
\bibitem{26} Cowling T G, 1941, MNRAS {\bf 101}, 367
\bibitem{27} Parker E N, 1966, ApJ {\bf 145}, 811
\bibitem{28} Krall N A and Trivielpiece A W, 1973, {\it Principles of plasma
physics}, McGraw-Hill, New York
\bibitem{29} Jackson J, 1972, Proc R Soc London {\bf A 200}, 248
\bibitem{30} Durrer R, Ferreira P G and Kahniashvil T, 2001, Phys Rev {\bf D 61},
43001
\bibitem{31} Vlahos L, Tsagas C G and Papadopoulos D B, 2005, ApJ Lett {\it in
press}
\bibitem{32} Tsagas C G and Maartens R, 2000, Phys Rev {\bf D 61}, 83519.
\bibitem{33} Parker E N, 1993, ApJ {\bf 414}, 389
\bibitem{34} Galeev A A and Sagdev R Z, 1984, {\it Basic Plasma Physics Vol 2}, A A
Galeev and R N Sudan (Eds), North Holand, Amsterdam
\bibitem{35} Hamberger S M and Friedmann M, 1968, Phys Rev Lett {\bf 21}, 674
\bibitem{36} Kronberg P P, Colgate S A, Li H and Dufton Q W, 2004, ApJ {\bf 604}, L77
\bibitem{37} Brandenburg A, Enqvist K and Olesen P, 1996, Phys Rev {\bf D
54}, 1291

\end{thebibliography}
\end{document}